# Dynamic Structural Recovery Parameters Enhance Prediction of Visual Outcomes After Macular Hole Surgery

Yinzheng Zhao[1], Zhihao Zhao[2], Rundong Jiang[3], Louisa Sackewitz[1], Quanmin Liang[4], Mathias Maier[1], Daniel Zapp[1], Peter Charbel Issa[1], Mohammad Ali Nasseri[1,5,6]

[1] Klinik und Poliklinik für Augenheilkunde, Technische Universität München, München, Germany
[2] TUM School of Computation, Information and Technology (CIT), Technische Universität München, München, Germany
[3] ShenShan Medical Center, Memorial Hospital of Sun Yat-Sen University, Guangzhou, China
[4] School of Computer Science and Engineering, Sun Yat-sen University, Guangzhou, China
[5] Department of Biomedical Engineering, University of Alberta, Alberta, Canada
[6] Zhongshan Ophthalmic Center, Sun Yat-sen University, Guangzhou, China

**Corresponding Author**:
Prof. M. Ali Nasseri
Klinik und Poliklinik für Augenheilkunde
Technische Universität München
Ismaninger Str. 22
81675 München, Germany
Email address: ali.nasseri@mri.tum.de

**Word Count**: 3,368

**Number of Figures**: 4   **Number of Tables**: 3   **Number of Supplementary Files**: 0

**Funding**: Supported by Bavarian Research Alliance grant funded by the Bavarian government (Germany; grant number AZ-1503-21)

**Commercial Relationships Disclosure**: [Y.Z. Zhao, None; Z.H. Zhao, None; R.D. Jiang, None; L.S, None; Q.M. Liang, None; M. Maier, None; D. Zapp, None; P.C. Issa, None, M.A. Nasseri, None]

**Running Head**: OCT Dynamics for Vision Prediction

**Abstract**

**Purpose:** To introduce novel dynamic structural parameters and evaluate their integration within a multimodal deep learning (DL) framework for predicting postoperative visual recovery in idiopathic full-thickness macular hole (iFTMH) patients.

**Methods:** We utilized a publicly available longitudinal OCT dataset at five stages (preoperative, 2 weeks, 3 months, 6 months, and 12 months). A stage specific segmentation model delineated related structures, and an automated pipeline extracted quantitative, composite, qualitative, and dynamic features. Binary logistic regression models, constructed with and without dynamic parameters, assessed their incremental predictive value for best-corrected visual acuity (BCVA). A multimodal DL model combining clinical variables, OCT-derived features, and raw OCT images was developed and benchmarked against regression models.

**Results:** The segmentation model achieved high accuracy across all timepoints (mean Dice $\geqslant$ 0.89). Univariate and multivariate analyses identified base diameter, ellipsoid zone integrity, and macular hole area as significant BCVA predictors ($P < 0.05$). Incorporating dynamic recovery rates consistently improved logistic regression AUC, especially at the 3-month follow-up. The multimodal DL model outperformed logistic regression, yielding higher AUCs and overall accuracy at each stage. The difference is as high as 0.12, demonstrating the complementary value of raw image volume and dynamic parameters.

**Conclusions:** Integrating dynamic parameters into the multimodal DL model significantly enhances the accuracy of predictions. This fully automated process therefore represents a promising clinical decision support tool for personalized postoperative management in macular hole surgery.

**Translational Relevance:** The integration of dynamic OCT-derived structural parameters into the multimodal DL framework enables personalized prediction of visual outcomes after macular hole surgery.

**Introduction**

Idiopathic Full-Thickness Macular Hole (iFTMH) is a retinal disorder characterized by a full-thickness defect in the central macula.[1] Clinically, it typically presents with symptoms including decreased central visual acuity, metamorphopsia, and central scotoma.[2-4] Despite the markedly increased hole closure rates achieved by surgical techniques, such as pars plana vitrectomy combined with internal limiting membrane peeling, functional visual recovery still exhibits substantial inter-individual variability.[5-8] This discrepancy between anatomical repair and visual outcome highlights the complexity of foveal remodeling and the need for deeper understanding of structural-functional relationships during postoperative recovery.

Current research primarily focuses on analyzing various morphological parameters extracted from OCT images using different software tools to assess their association with postoperative visual recovery.[9-13] However, most studies heavily rely on static morphological measurements or qualitative assessments at limited time points, which often fail to capture the dynamic progression of retinal healing following surgery.[14, 15] Second, while deep learning techniques have been increasingly applied to OCT imaging in other retinal diseases, their application in iFTMH remains relatively nascent,[16-18] particularly in integrative disease modeling across multiple time points. Moreover, the majority of studies remain limited by manual or semi-automated workflows.[15, 19-21] These studies provide limited insight into the dynamic processes of foveal tissue recovery and the temporal evolution of key retinal structures after surgery.

To address these gaps, we constructed a longitudinal analytical framework that leverages automated OCT segmentation and quantification across multiple postoperative stages (Figure 1). Our system not only captures classic structural features but also introduces a novel category of dynamic parameters that reflect the speed and trajectory of tissue recovery over time. By integrating clinical data, quantitative morphological metrics, and time-series indicators, we aim to provide a more nuanced understanding of how different retinal structures contribute to postoperative visual rehabilitation.

Our work offers three key contributions. First, we propose a longitudinal OCT segmentation model tailored to the structural variability of preoperative and postoperative stages. Second, we introduce the dynamic parameters as a novel category of descriptors. Third, we establish a multimodal prediction model that combines clinical, structural, and longitudinal sequential images containing temporal information to estimate postoperative visual acuity. This integrative approach may provide new insights into the changes in macular tissue repair of the retina and offer practical tools for individualized prognosis after macular hole surgery.

**Methods**

*Segmentation Model for Structural Feature Delineation*

To accurately quantify the morphological characteristics of iFTMH and their dynamic changes after surgery, we first employed nnUNet[22] for the segmentation of pathological regions in OCT images. The dataset was stratified according to clinical follow-up stages into five subsets, preoperative, and postoperative at 2 weeks, 3 months, 6 months, and 12 months. Each examination included two orthogonal B-scans (horizontal and vertical), centered on the fovea. According to the significant anatomical variability across different stages, the model was specifically trained to recognize distinct pathological features present at each time point. Preoperative images (Figure 2) primarily involved annotations of macular holes, pseudocysts, epiretinal membranes (ERM), vitreomacular traction (VMT), posterior vitreous detachment (PVD), external limiting membrane (ELM), ellipsoid zone (EZ), and retinal pigment epithelium (RPE). In contrast, postoperative scans focused on monitoring the closure of the macular hole, resolution of pseudocysts, and the restoration of outer retinal structures.

Two junior retinal specialists independently annotated images as the GroundTruth, and a senior retinal specialist reviewed the labels to ensure annotation consistency and accuracy. We manually annotated 150 images at each stage. In total, across five stages, we had 750 images, with 500 used for training and 250 for testing. The Dice similarity coefficient was employed to evaluate the segmentation performance. During the testing phase, when the loss function stabilized, the average Dice reached 0.862, and we adopted this model as the foundation of our automatic segmentation model. Subsequently, we performed fully automatic segmentation on a total of 2,591 images across all five stages. The segmented pathological regions served as the basis for quantitative feature extraction in subsequent analyses, enabling precise evaluation of the structural recovery process over time.

*Automated Feature Quantification and Dynamic Parameter Derivation*

Based on the segmented OCT images, we implemented a fully automated pipeline to extract and quantify key morphological features. All extracted parameters were classified into four main categories, including quantitative parameters, composite parameters, qualitative parameters, and dynamic parameters. Parameter sets were collectively termed "Values" and were integrated into the multimodal modeling framework for further predictive analyses.

Quantitative parameters refer to direct morphometric measurements derived from the segmented lesion regions. These include the minimum linear diameter (MLD) and base diameter (BD) of the macular hole, the shortest vertical distance (e) between MLD and BD, the height of the macular hole (height), the area of pseudocysts, the area of the macular hole, and the length of outer retinal defects such as disruptions in the ELM and the EZ.

Composite parameters were calculated by deriving ratios or indices from these basic measurements to enhance their clinical interpretability and correlation with visual outcomes. Drawing upon prior literatures and clinical convention, we included the macular hole index (MHI, defined as height divided by BD), the traction hole index (THI, height divided by MLD), and the diameter hole index (DHI, MLD divided by BD).[23] Additionally, we computed the area ratio between the macular hole and surrounding pseudocysts.[20] Qualitative parameters included the presence or absence of epiretinal membrane (ERM) and the identification of a tractional space between the ERM and the inner retinal surface.

Importantly, to capture the dynamic healing process, we introduced a novel category of dynamic parameters. These parameters focused on the temporal evolution of key features, specifically the macular hole, pseudocysts, the ELM, and the EZ. For metrics such as macular hole area and outer retinal defect length, recovery rate was defined as the ratio between the initial lesion size on preoperative OCT and the time point at which the lesion was observed to be fully resolved. With permitted data, we further incorporated a shape-weighted factor to account for pre-resolution morphological complexity, which was included in the dynamic parameter computation through a coefficient-weighted adjustment model.

*Multimodal Deep Learning Prediction Model*

To comprehensively predict postoperative visual recovery, we constructed a multimodal DL that jointly incorporates three distinct data modalities, including CD, image derived feature data (Values), and OCT image data. This framework aims to leverage the complementary strengths of each modality, structured clinical variables, high-dimensional quantitative features extracted via our automated analysis pipeline, and the rich spatial information retained in raw OCT images, to achieve robust and individualized prognostic modeling. Our model structure is shown in Figure

3. Our model accepts multiple inputs, with images as the primary input. On this basis, we use cross attention[24] to integrate clinical numerical information and pathological parameter numerical information. Specifically, we first learn a CLIP model using ViT as the image encoder on a public OCT dataset containing 30,000 image-text pairs. Then, we take the image encoder of the CLIP model as the backbone of our prediction model for feature extraction to obtain Fi. For clinical data and parameter numerical data, we first connect them separately to multilayer perceptron (MLP) structures, mapping two numerical vectors to the same feature space via MLPs to obtain Fc and Fv, respectively. Next, we fuse the features of different modalities through cross attention by connecting Fc and Fv with Fi via CrossAttention, leveraging its cross-modal fusion capabilities. Finally, we concatenate the features and input them into two MLP structures to map data features to classification heads, generating probability predictions through the classification heads.

The purpose of the model is to predict postoperative BCVA improvement based on preoperative parameters. Evaluations are conducted at four clinically relevant time points: 2 weeks, 3 months, 6 months, and 12 months post-surgery. A difference of 20 or more in ETDRS scores between each period and preoperatively indicates significant postoperative BCVA recovery (Superior). Typically, a threshold is set at 15; however, to raise the requirements for model prediction in this study, we set it at 20.

To assess the added value of multimodal integration, we compared the performance of our model against traditional logistic regression classifiers using single and bimodal input types. Model performance was evaluated using standard metrics including accuracy, sensitivity, specificity, and area under the receiver operating characteristic curve (AUC). We split the dataset into training and test sets with a ratio of 8:2, conducted five-fold cross-validation on the training set, and finally evaluated on the test set.

*Logistic Regression Model and Dataset*

To benchmark the performance of our multimodal deep learning model, we additionally constructed a traditional binary logistic regression model using two structured data modalities: clinical data and Values data derived from our fully automated feature analysis system. This model served as a baseline to evaluate the incremental predictive value of incorporating raw OCT imaging data and deep neural architectures.

Data cleaning and preprocessing were conducted iteratively using both SPSS (IBM SPSS Statistics, Version 27.0) and Python (v3.10). For missing data handling, entries with <10% missing values were imputed using the mean value of the corresponding variable. Variables with greater proportions of missingness were excluded from analysis to reduce imputation bias. Normality tests were conducted on all continuous variables using the Shapiro-Wilk test, and Pearson or Spearman correlation analysis was applied depending on distribution characteristics. Multicollinearity among predictors was assessed using the variance inflation factor (VIF), and variables with VIF > 5 were excluded to maintain model stability.

The logistic regression model was developed to predict different periods postoperative visual recovery, defined as a binary outcome based on BCVA (Superior vs. not Superior). Feature selection was carried out through univariate logistic regression screening ($p < 0.10$). Model performance was assessed via receiver operating characteristic (ROC) analysis, with AUC, sensitivity, specificity, and overall accuracy as evaluation metrics.

All OCT images and corresponding clinical metadata used in this study were derived from a publicly available dataset, provided by the Ophthalmic Imaging Research Group at CHU de Québec, and accessible at *https://www.kaggle.com/datasets/mathieugodbout/oct-postsurgery-visual-improvement*. [11]

**Results**

*Segmentation Model Performance*

We first evaluated the performance of our segmentation model. We used DICE, IOU, and ROC_AUC as metrics to assess the accuracy of our segmentation (Table 1). Across multiple pathological categories, the average performance achieved with these metrics was as follows: DICE 0.862, IOU 0.774, and ROC_AUC 0.912. We further assessed the model's segmentation precision for each labeled anatomical or pathological structure. Structures with well-defined morphological boundaries, such as the macular hole, RPE, EZ, and ELM, exhibited higher Dice scores (MH: 0.945; EZ: 0.875; RPE: 0.906; ELM: 0.868). Structures exhibiting greater inter-patient variability or temporal change, such as ERM and VMT, yielded relatively lower but acceptable Dice scores (ERM: 0.805; PVD: 0.853), likely reflecting biological variability and transient presence at different stages.

*Logistic Regression Model of Feature Parameters*

To evaluate the predictive value of clinical and imaging-derived features on postoperative visual outcomes, we first performed univariate correlation analyses between each parameter and BCVA at three key follow-up timepoints, postoperative week 2, month 3, and month 12. Disease duration, BD, MLD, e, MHI, defect length of EZ and ELM, were significant ($P < 0.05$). Subsequently, we constructed binary logistic regression models to classify postoperative visual recovery status (Superior vs. not Superior) based on the aggregated feature set. In the initial model, quantitative, qualitative, and composite parameters were included as predictors after screening. To further assess the added value of our newly introduced dynamic parameters, we extended the model to include these temporal features. As shown in Table 2, the inclusion of diseases recovery rates led to moderate but consistent improvements in model performance across all timepoints.

In the final binary logistic regression model constructed (Table 3), we identified multiple parameters that were significant and had predictive value for postoperative visual recovery at different stages. At two weeks postoperatively, disease duration (OR =0.884, P =0.006), macular hole area (OR =1.001, P =0.003), and BD (OR =0.968, P =0.024) were all significantly associated with early BCVA improvement. For each additional 1.0 μm increase in BD, the short-term likelihood of visual improvement decreased by approximately 3.2%. In the three-month postoperative model, for each additional 1.0 μm increase in MLD, the likelihood of visual improvement decreased by approximately 5.2%, consistent with BD findings, indicating a significant negative impact of hole size on intermediate-term outcomes. Additionally, dynamic parameters such as EZ recovery rate (OR =1.298, P =0.048) and macular hole recovery rate (OR =0.993, P =0.049) were included in the final model. For each percentage point increase in EZ recovery rate, the likelihood of visual improvement increased by approximately 29.8%, highlighting the unique value of dynamic parameters in intermediate-term prognosis assessment. At 12 months postoperatively, EZ defect length (OR =1.017, P =0.048) and cystoid area (OR =0.999, P =0.026) still showed statistical significance, though with relatively smaller effect sizes. Overall, these results suggest that postoperative visual recovery at different time points is influenced by multiple factors, particularly in the intermediate term where a combination of structural and dynamic parameters enhances the accuracy of prognosis evaluation.

*Enhanced Multimodal Deep Learning Outperforms Traditional Logistic Models*

To evaluate the predictive performance of different modeling strategies for postoperative BCVA improvement, we compared deep learning (DL) and logistic regression (LG) models across various data modality combinations, with or without (w/o) dynamic parameters (DP). As shown in Figure 4, the multimodal DL model combining clinical data (CD), structural features (Values, including DP), and raw OCT images consistently outperformed all other configurations across all follow-up timepoints.

Notably, at 2 weeks (Figure 4A), this tri-modal DL system achieved an AUC of 0.94, markedly higher than the image-only DL (0.91) and the LG (CD and Values, 0.87), underscoring the benefit of early incorporation of dynamic healing cues. By 3 months (Figure 4B), the advantage persisted. The full DL model maintained an AUC of 0.90 versus 0.85 for the best-performing regression model, highlighting its robustness in the critical mid-term recovery window. At 6 and 12 months (Figures 4C and 4D), although overall discrimination modestly declined, reflecting increased heterogeneity in long-term healing, the multimodal DL approach still outperformed all reduced models (AUCs of 0.91 at 6 months and 0.89 at 12 months). In contrast, excluding DP from either framework generally reduced AUC by 0.03–0.04. These findings demonstrate that integrating temporal dynamics with imaging and clinical information yields a stable, high-fidelity predictor of visual outcomes, particularly valuable in the early and mid-term postoperative phases.

Ablation analysis further demonstrated the additive value of each data modality. Models trained with only text parameters (CD and Values) or only images showed limited performance, while the combination of all three modalities produced the highest predictive accuracy and AUC. These findings suggest that deep learning can effectively synthesize complementary information from different data domains, including image derived textures, spatial relationships, and clinically meaningful morphometric trends, to enhance functional outcome prediction.

**Discussion**

The structural determinants of visual recovery following iFTMH surgery have been extensively investigated, with consistent evidence supporting the prognostic value of parameters such as MLD, BD, and integrity of the ELM.[19, 25-27] In line with these prior findings, our study confirmed the predictive relevance of these anatomical features at multiple postoperative stages, strengthening the robustness and clinical validity of our automated analysis pipeline. By leveraging longitudinal OCT data and standard clinical information, we established a consistent structure-function association across both early and late recovery periods.

The innovation of this study is the introduction of dynamic parameters that quantify the rate and extent of structural recovery over time. Rather than relying solely on static morphological metrics, we characterized how features such as macular hole area, pseudocyst presence, and outer retinal defect length evolved postoperatively. These dynamic indicators provided an additional dimension to the understanding of postoperative healing, especially at the 3-month timepoint, where recovery velocity of the EZ was significantly associated with visual outcomes, suggesting that structural recovery may begin earlier than typically appreciated. Mechanistically, the dynamic recovery rate of the ellipsoid zone (EZ) on OCT likely reflects the reconstitution and realignment of photoreceptor outer segments, as this hyperreflective band corresponds to the mitochondria-rich inner segment ellipsoids critical for phototransduction.[28, 29] Likewise, restoration of the external limiting membrane (ELM) indicates reformation of Müller cell–photoreceptor junctions, underscoring its role as an early scaffold for photoreceptor regeneration and a predictor of subsequent functional recovery.[30] Clinically, such temporal descriptors may assist in stratifying patients who are likely to benefit from closer monitoring or tailored postoperative care.

The superior performance of the multimodal deep learning model in predicting visual recovery further supports the value of integrating diverse information sources. Beyond predefined morphometric features and clinical variables, the inclusion of raw OCT images allowed the model to access richer spatial and textural cues, potentially capturing subtle signs of tissue remodeling or microarchitectural distortion not easily measurable. Our model demonstrated consistently superior predictive performance compared to traditional logistic regression approaches, especially in identifying patients with suboptimal visual outcomes. Ablation studies revealed that the inclusion of dynamic parameters and raw OCT image data contributed significantly to model accuracy, underscoring the added prognostic value of spatial-temporal imaging features beyond conventional clinical and morphometric descriptors. The consistent advantage of tri-modal input over unimodal or bimodal configurations suggests that these data types are complementary, and their integration can lead to more reliable and individualized prognostication. With further validation, such models may be embedded into clinical workflows to assist in outcome counseling and adaptive treatment planning.

This study has several limitations. The data were obtained from a single publicly available dataset with moderate sample size, limiting the model's external generalizability. While the dynamic parameters offered new insight, their calculation was based on discrete follow-up intervals, and may not fully capture the continuous nature of retinal healing. Additionally, BCVA was the sole outcome variable, and future studies incorporating functional vision metrics such as contrast sensitivity or microperimetry would provide a more comprehensive assessment. Despite these limitations, our study demonstrates the potential of dynamic structural descriptors and multimodal modeling in improving postoperative outcome prediction in macular hole surgery.

**Conclusion**

In this study, we developed a fully automated pipeline for longitudinal OCT analysis and postoperative outcome prediction in idiopathic full-thickness macular hole. By introducing dynamic structural parameters and integrating clinical, quantitative, and imaging data within a multimodal framework, we demonstrated improved prediction of visual recovery at multiple timepoints. These findings highlight the potential value of incorporating temporal structural metrics and deep learning–based models into individualized postoperative care strategies. Future work with larger, multi-center datasets will be essential to validate these findings and support clinical translation.

**Data Availability Statement**

The OCT imaging data and associated clinical variables used in this study were obtained from a publicly available dataset originally published by Godbout et al. in their work titled *"Predicting Visual Improvement after Macular Hole Surgery: A Cautionary Tale on Deep Learning with Very Limited Data"*. The dataset can be accessed and downloaded from Kaggle via the following link: https://www.kaggle.com/datasets/mathieugodbout/oct-postsurgery-visual-improvement. All data were used in accordance with the usage terms specified by the dataset providers.

**Acknowledgements**

Supported by Bavarian Research Alliance grant funded by the Bavarian government (Germany; grant number AZ-1503-21). The Bavarian government had no role in the design or conduct of the study.

**Tables**

**Table 1.** The segmentation accuracy of the model across multiple pathological categories.

| Class | Dice | IOU | Accuracy | F1-Score | ROC_AUC |
|---|---|---|---|---|---|
| ELM | 0.8676 | 0.7681 | 0.9984 | 0.8676 | 0.9269 |
| EZ | 0.8749 | 0.7795 | 0.9985 | 0.8749 | 0.9151 |
| RPE | 0.9056 | 0.8288 | 0.9977 | 0.9056 | 0.9539 |
| Macular Hole | 0.9445 | 0.9057 | 0.9987 | 0.9445 | 0.9864 |
| Pseudocysts | 0.8904 | 0.8096 | 0.9987 | 0.8904 | 0.9541 |
| PVD | 0.8528 | 0.7577 | 0.9974 | 0.8528 | 0.9618 |
| VMT | 0.8546 | 0.7563 | 0.9974 | 0.8546 | 0.9525 |
| Space | 0.7595 | 0.6617 | 0.9971 | 0.7595 | 0.6666 |
| ERM | 0.8048 | 0.6958 | 0.9989 | 0.8048 | 0.8907 |
| Mean | 0.8616 | 0.7737 | 0.9981 | 0.8616 | 0.9120 |

**Table 2.** The effect of a logistic regression model on predicting postoperative BCVA status at different time points including or excluding dynamic parameters.

|  | PO 2 weeks | PO 3 months | PO 6 months | PO 12 months |
|---|---|---|---|---|
| Baseline | 79.50% | 69.10% | 52.50% | 54.70% |
| Prediction w/o Dynamic Parameters | 86.20% | 81.80% | 81.00% | 78.60% |
| Prediction with Dynamic Parameters | 87.60% | 84.30% | 84.10% | 80.20% |
| **P value** | P＜0.001 | P＜0.001 | P＜0.001 | P＜0.001 |
| Nagelkerke $R^2$ (with DP) | 0.572 | 0.542 | 0.688 | 0.586 |

*$R^2 > 0.3$ means significant; PO means post-operative; DP means dynamic parameters*

**Table 3.** Binary classification logistic regression model predicts the results of postoperative short-term and long-term BCVA improvement status.

| | B | S.E. | Wald | Sig. (P value) | Exp (B) (OR) | Exp (B) 95% CI Lower Limits | Exp (B) 95% CI Upper Limits |
|---|---|---|---|---|---|---|---|
| Post operative 2 weeks | | | | | | | |
| Duration (days) | -0.124 | 0.045 | 7.563 | 0.006 | 0.884 | 0.809 | 0.965 |
| Macular Hole Area | 0.001 | 0.000 | 8.593 | 0.003 | 1.001 | 1.000 | 1.001 |
| BD | -0.033 | 0.014 | 5.112 | 0.024 | 0.968 | 0.941 | 0.996 |
| Post operative 3 months | | | | | | | |
| EZ Defect Length | -0.012 | 0.005 | 6.248 | 0.012 | 0.988 | 0.979 | 0.997 |
| MLD | -0.053 | 0.023 | 5.477 | 0.019 | 0.948 | 0.906 | 0.991 |
| Recovery Rate (Macular Hole) | -0.007 | 0.004 | 3.660 | 0.050 | 0.993 | 0.986 | 1.000 |
| Recovery Rate (EZ) | 0.261 | 0.133 | 3.855 | 0.048 | 1.298 | 1.001 | 1.685 |
| Post operative 12 months | | | | | | | |
| EZ Defect Length | 0.017 | 0.009 | 3.903 | 0.048 | 1.017 | 1.000 | 1.034 |
| Pseudocysts Area | -0.001 | 0.000 | 4.968 | 0.026 | 0.999 | 0.999 | 1.000 |

*95% CI: 95% Confidence Interval；S.E.: Standard Error；P<0.05 means significant*

**Figure Captions**

**Figure 1.** Workflow of the multimodal OCT-based prediction process. The pipeline begins with data acquisition of a public longitudinal OCT and clinical dataset. A longitudinal segmentation model then delineates macular hole–related structures at each follow-up stage. Segmentation outputs feed into a fully automated feature extraction step, yielding quantitative, composite, qualitative, and novel dynamic parameters. These features, along with raw OCT images and clinical variables, serve as inputs to both logistic regression (with/without dynamic parameters) and multimodal deep learning models. Final evaluation compares model discrimination (AUC, ACC, P value) and includes ablation studies of data modalities to demonstrate the added value of dynamic parameters and raw image information.

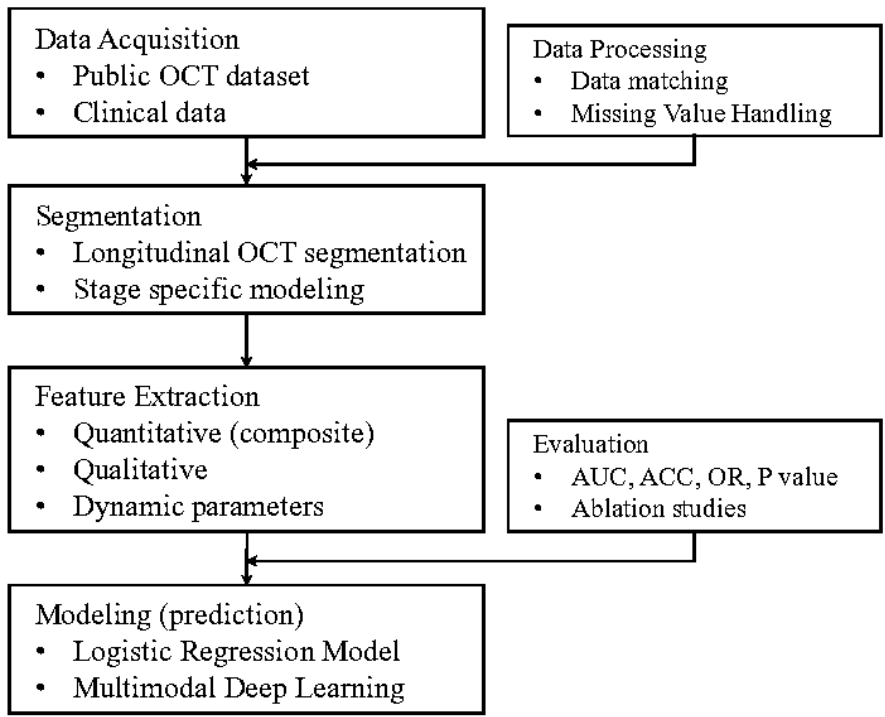

**Figure 2.** Segmentation models can perform precise segmentation on different disease characteristics. Different colors respectively represent macular hole, pseudocyst, ERM, space, VMT, PVD, ELM, EZ and RPE.

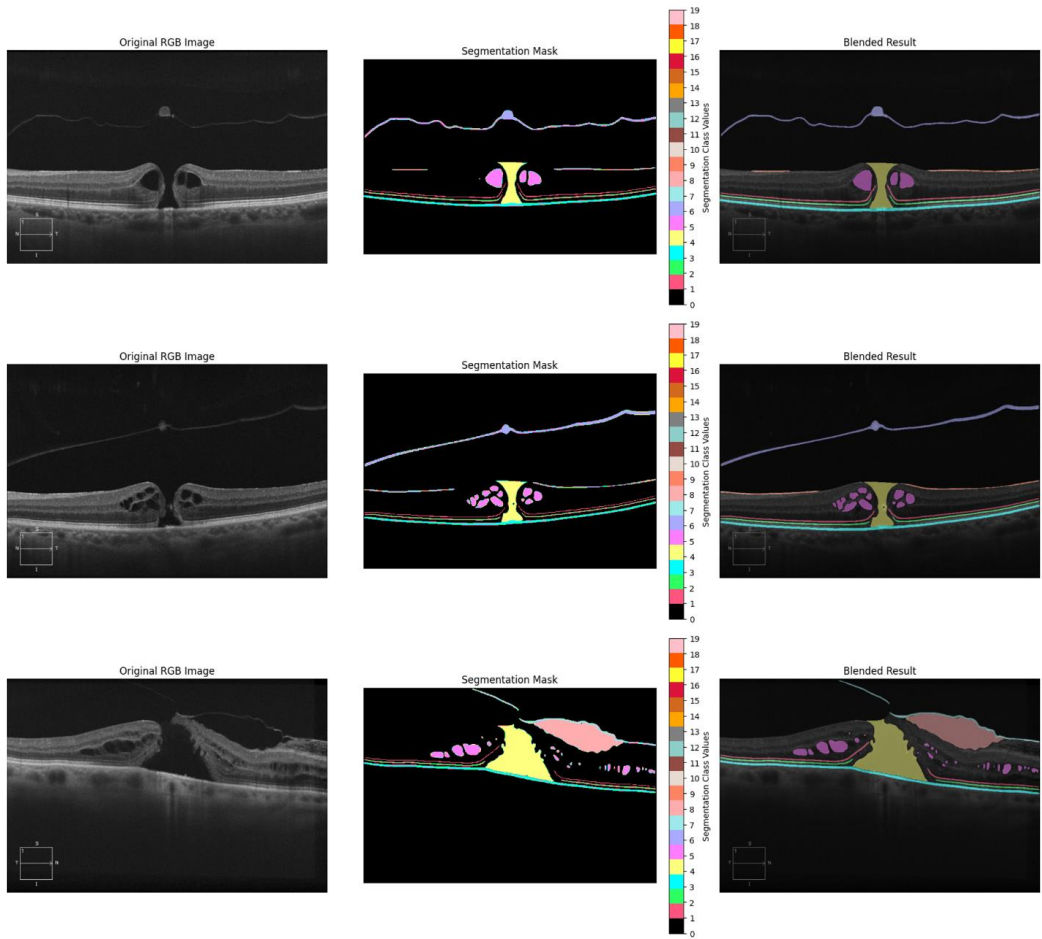

**Figure 3.** Architecture of the multimodal deep learning model for predicting postoperative visual recovery. The model integrates three data modalities: structured clinical data (Vc), quantitative feature vectors extracted from OCT segmentation (Vv), and raw OCT images (Vi). Clinical and feature vectors are encoded via multilayer perceptrons (MLPs) to obtain Fc and Fv, respectively, while OCT images are processed through a pre-trained ViT based image encoder (Fi). Cross-attention modules are employed to fuse numerical feature vectors with image-derived representations. The concatenated features are passed through a final encoder and classification head to generate probability outputs for visual improvement. A schematic of the cross-attention mechanism is shown in the lower right.

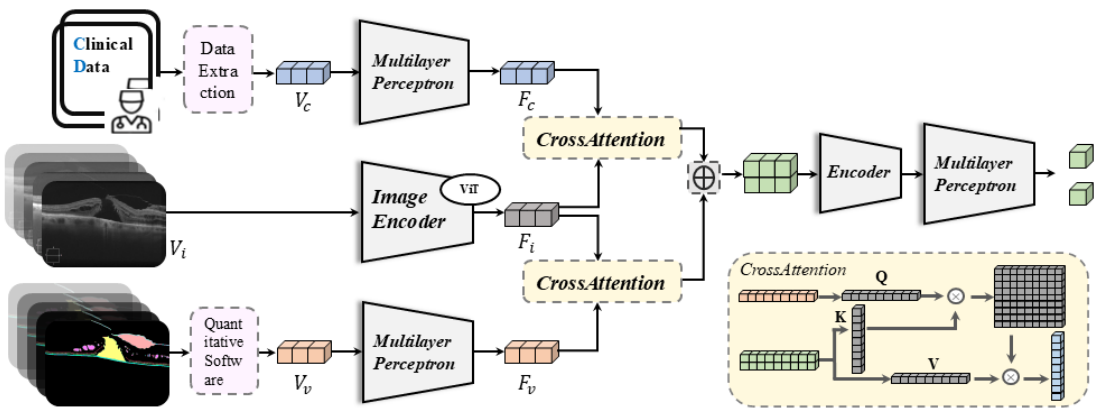

**Figure 4.** Receiver operating characteristic (ROC) curves comparing the predictive performance of deep learning (DL) and logistic regression (LG) models across different data modality combinations, with or without dynamic parameters (DP), for postoperative best-corrected visual acuity (BCVA) improvement at four timepoints, (A) 2 weeks, (B) 3 months, (C) 6 months, and (D) 12 months after surgery. CD means clinical data; Values means extracted structural features. Across all stages, the DL model integrating CD, Values (including DP), and images achieved the highest AUC. Exclusion of DP led to a consistent decline in performance, particularly in mid-term prediction. The results highlight the additive value of dynamic structural parameters and image-level representations in modeling visual recovery.

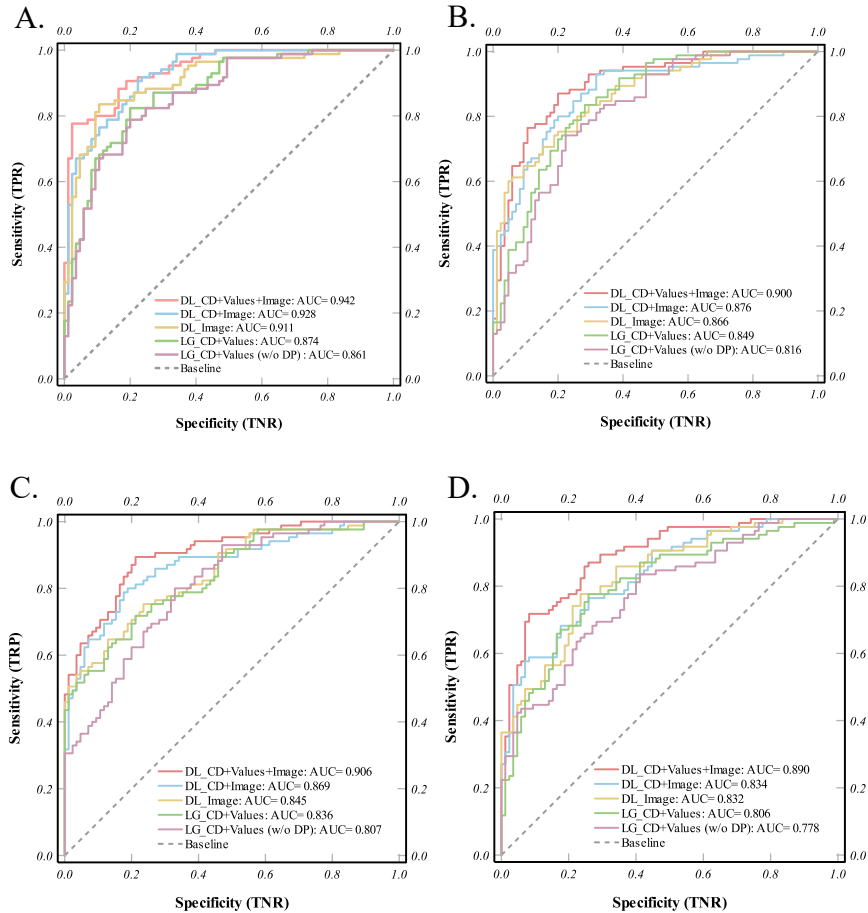